\documentclass[aps,prx,10pt,twocolumn,longbibliography,superscriptaddress,floatfix]{revtex4-2}
\usepackage{graphicx}
\usepackage{dcolumn}
\usepackage{bm}
\usepackage{hyperref}
\usepackage{gensymb}
\hypersetup{
    colorlinks=true,
    urlcolor= blue,
    citecolor=blue,
linkcolor= blue}
\usepackage{amsmath}
\usepackage[skip=5pt, indent=10pt]{parskip}

\begin{document}

\preprint{APS/123-QED}

\title{Andreev processes in mesoscopic multi-terminal graphene Josephson junctions}

\author{Fan Zhang}
\affiliation{Department of Physics, The Pennsylvania State University, University Park, PA 16802, USA}
\author{Asmaul Smitha Rashid}
\affiliation{Department of Electrical Engineering, The Pennsylvania State University, University Park, PA 16802, USA}
\author{Mostafa Tanhayi Ahari}
\affiliation{Materials Research Laboratory, The Grainger College of Engineering, University of Illinois, Urbana-Champaign, IL 
61801, USA}
\author{Wei Zhang}
\affiliation{Department of Electrical Engineering, The Pennsylvania State University, University Park, PA 16802, USA}
\author{Krishnan Mekkanamkulam Ananthanarayanan}
\affiliation{Department of Electrical Engineering, The Pennsylvania State University, University Park, PA 16802, USA}
\author{Run Xiao}
\affiliation{Department of Physics, The Pennsylvania State University, University Park, PA 16802, USA}
\author{George J. de Coster}
\affiliation{DEVCOM Army Research Laboratory, 2800 Powder Mill Rd, Adelphi, MD, 20783, USA}
\author{Matthew J. Gilbert}
\affiliation{Department of Electrical Engineering, University of Illinois, Urbana-Champaign, IL 61801, USA}
\affiliation{Materials Research Laboratory, The Grainger College of Engineering, University of Illinois, Urbana-Champaign, IL 61801, USA}

\author{Nitin Samarth}%
\thanks{Corresponding author: nsamarth@psu.edu}
 \affiliation{Department of Physics, The Pennsylvania State University, University Park, PA 16802, USA}

\author{Morteza Kayyalha}%
\thanks{Corresponding author: mzk463@psu.edu}
 \affiliation{Department of Electrical Engineering, The Pennsylvania State University, University Park, PA 16802, USA}

\date{\today}

\begin{abstract}
There is growing interest in using multi-terminal Josephson junctions (MTJJs) as a platform to artificially emulate topological phases and to investigate complex superconducting mechanisms such as quartet and multiplet Cooper pairings. Current experimental signatures in MTJJs have led to conflicting interpretations of the salient features. In this work, we report a collaborative experimental and theoretical investigation of graphene-based four-terminal Josephson junctions. We observe resonant features in the differential resistance maps that resemble those ascribed to multiplet Cooper pairings. To understand these features, we model our junctions using a circuit network of coupled two-terminal resistively and capacitively shunted junctions (RCSJs). Under appropriate bias current, the model predicts that supercurrent flow between two terminals in a four-terminal geometry may be represented as a sinusoidal function of a weighted sum of the superconducting phases. We find that the resonant features generated by the RCSJ model are insensitive to the diffusive or ballistic form of the current-phase relation and junction transparency. Our study suggests that differential resistance measurements alone are insufficient to conclusively distinguish resonant Andreev reflection processes from semi-classical circuit-network effects.
\end{abstract}

\maketitle

\section{Introduction}
The Josephson effect is a centerpiece of many quantum device applications, including superconducting quantum interference devices (SQUIDs) and superconducting qubits \cite{kjaergaard2020superconducting,de2021materials}. Increasing the number of superconducting terminals beyond the typical two terminals in Josephson junctions (JJs) leads to non-local coupling of superconducting order parameters through a common scattering region. This non-local coupling has been predicted to lead to quartet Cooper pairings \cite{PhysRevLett.106.257005} and macroscopic multi-channel effects such as phase drag and magnetic flux transfer \cite{de1995multi,amin2001mesoscopic,amin2002dc}; it may also be used to form a superconducting phase qubit \cite{amin2002multi}. More recently, multi-terminal Josephson junctions (MTJJs) have been proposed as a platform to emulate topological phases in artificial dimensions \cite{van2014single,yokoyama2015singularities,riwar2016multi,eriksson2017topological,meyer2017nontrivial,xie2018weyl,xie2017topological,xie2021non,gavensky2019topological,xie2019topological,erdmanis2018weyl,repin2019topological,huang2019topology,strambini2016omega}. In MTJJs with $n$ superconducting terminals, the energy of the Andreev bound states is a function of ($n-1$) independent phases. In this context, the phase differences between superconducting terminals are treated as quasi-momenta of a crystal forming a Brillouin zone in ($n-1$) dimensions. The resultant band structure may display topological properties such as Weyl singularities \cite{riwar2016multi,eriksson2017topological}. While providing strong motivation for studying the multi-terminal Josephson effect, the exploration of MTJJs as a platform for engineering artificial topological systems is still nascent.

MTJJs have been experimentally explored in various materials platforms including graphene/MoRe \cite{draelos2019supercurrent,huang2022evidence,arnault2022dynamical} and InAs/Al \cite{cohen2018nonlocal,pankratova2020multiterminal,graziano2022selective,graziano2020transport,gupta2022superconducting}. These experiments focused on the gate and magnetic field dependence of the supercurrent flow between adjacent and nonadjacent superconducting terminals \cite{graziano2022selective}. Additionally, these experiments studied the non-trivial geometric response of the critical current contour (CCC), a generic characteristic defining the region in which all the superconducting terminals are at zero voltage \cite{pankratova2020multiterminal}. Recent studies have also reported signatures of quartet pairings \cite{cohen2018nonlocal,pfeffer2014subgap,huang2022evidence}, arising from crossed Andreev reflection processes. However, studies of three-terminal Josephson junctions \cite{arnault2022dynamical} and a network of tunnel junctions \cite{melo2022multiplet} argue that the resonant features may also arise from circuit-network effects.

In this paper, we use a coordinated experimental and theoretical approach to critically understand and examine the transport manifestations of Andreev processes in four-terminal JJs fabricated in hBN-encapsulated graphene heterostructures. We use device geometries wherein the superconducting terminals are connected via a common scattering region such that no individual two-terminal JJ is formed between the adjacent superconductors (Fig.~\ref{FIG1-01_NEW.png}). Hence, our device geometry allows us to better explore phenomena that are due to non-local coupling of superconducting order parameters. We model our junctions using a circuit network of coupled resistively and capacitively shunted junctions (RCSJs). We show that the semi-classical RCSJ model reproduces the observed resonant features, which are similar in nature to those predicted for multiplet Cooper pairings \cite{pfeffer2014subgap,cohen2018nonlocal,huang2022evidence}. To elucidate the underlying mechanisms giving rise to these features, we calculate the contribution of the quasiparticle current which reveals the pair current contribution to the total current. We further consider materials-specific properties such as the Fermi surface geometry of the normal material and junction transparency by incorporating the relevant current-phase relation (CPR) into the RCSJ model. We show that while the Fermi surface geometry is primarily responsible for the shape of the CCC~\cite{pankratova2020multiterminal}, the resonant features outside of the CCC are robust to changes in the Fermi surface and junction transparency. Finally, our study demonstrates that circuit-network effects, predicted by the RCSJ model, lead to macroscopic signatures in differential resistance maps that are identical to those ascribed to distinctly quantum processes such as multiplet pairings.

The paper is organized as follows. In Sec.~\ref{device_fab}, we discuss the details of device fabrication. In Sec.~\ref{asymmetry}, we describe the transport data for the asymmetric device. We also theoretically analyze our junctions using an RCSJ model. In Sec.~\ref{symmetry}, we discuss the transport results in the symmetric device. We establish that multiple Andreev reflections (MARs) are responsible for the discrepancies between the experimental and theoretical data. In Secs.~\ref{ballistic} and \ref{phi_0}, we consider the effect of different CPRs ranging from ballistic to diffusive transport in the RCSJ model and analyze the resultant differential resistance maps. We conclude in Sec.~\ref{conclusion}. 

\section{Device fabrication}\label{device_fab}
\begin{figure}
\includegraphics{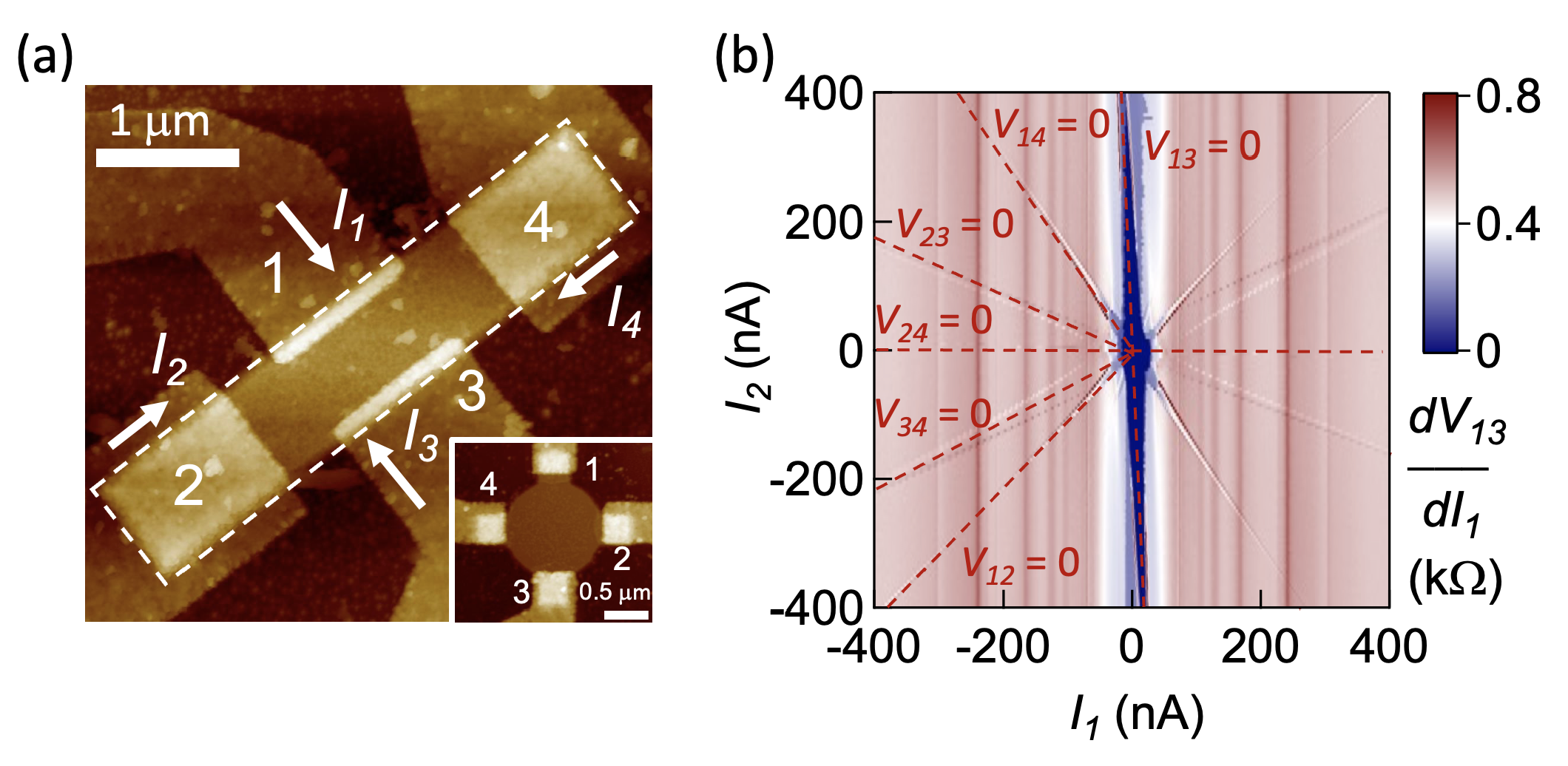}
\caption{\label{FIG1-01_NEW.png} (a) The AFM image of device A. Inset shows the AFM image of device B. The Josephson junctions are made of hBN-graphene-hBN (outlined by white dashed line) edge contacted with superconducting Al terminals. Arrows show the directions of bias currents. (b) Color map of the differential resistance ($dV_{13}/dI_{1}$) versus applied dc current biases $I_{1}=-I_{3}$ and $I_{2}=-I_{4}$, as indicated in panel (a) of device A. The differential
resistance is measured using a lock-in amplifier. Dashed red lines indicate superconducting branches corresponding to $V_{jk}$ = 0, where $j$ and $k$ are pairs of superconducting terminals as labeled in (b). Vertical lines in the map along the $I_{2}$ direction are multiple Andreev reflections corresponding to $eV_{13} = 2\Delta/n$ with $\Delta$ $\approx$ 169 $\mu$eV. All the measurements are performed at $V_g = 50$ V and $T = 12$ mK.}
\end{figure}

We assemble hBN/graphene/hBN van der Waals heterostructures using a standard dry transfer technique, followed by annealing in H$_2$/Ar gas at 350 $^{\circ}$C to remove polymer residues from the heterostructures \cite{wang2013one,purdie2018cleaning}. The heterostructures are then patterned with electron beam lithography, followed by dry etching (O$_2$/CHF$_3$), to define the junction area. Another e-beam lithography step is performed to define the contact patterns. Finally, Ti(10 nm)/Al(100 nm) is evaporated to create superconducting edge contacts. Atomic force microscope (AFM) images of our four-terminal JJs are shown in Fig.~\ref{FIG1-01_NEW.png}(a). Devices A and B are four-terminal asymmetric and symmetric junctions, respectively. The channel length of device A is $0.8$ $\mu$m and $3$ $\mu$m along $I_{13}$ and $I_{24}$ directions, respectively. Device B has a circular geometry with a diameter of $1.3$ $\mu$m. The Dirac point of device A and B is at $V_g = -4.5$ V and $V_g = -11.25$ V, respectively. In Fig.~\ref{FIG1-01_NEW.png}(a), we show the current directions used for the transport measurements. We set $I_{1}=-I_{3}$ and $I_{2}=-I_{4}$ in our experiments. We perform all of the measurements at $T = 12$ mK, if not otherwise specified. 

\section{Transport in the asymmetric device geometry}\label{asymmetry}
We now turn our attention to the transport characteristics of the multi-terminal junctions in device A. Fig.~\ref{FIG1-01_NEW.png}(b) shows a differential resistance $dV_{13}/dI_{1}$ map versus $I_{1}$ and $I_{2}$ measured at $V_g = 50$ V. The directions indicated by the red dashed lines are branches along the local minima of $dV_{13}/dI_1$ and satisfy $V_{jk} = 0$ conditions as labeled. These conditions correspond to supercurrent flow between terminals $j$ and $k$. The vertical lines in the map along the $I_{2}$ direction are signatures of MARs. Along these lines, $eV_{13} \approx 2\Delta/n$, where $e$ is the electron charge, $\Delta$ is the superconducting gap, and $n$ is an integer. We calculate an induced superconducting gap of $\Delta \approx$ 169 $\mu$eV by fitting $eV_{13}$ to $2/n$. The calculated gap is consistent with the gap we obtain from the temperature-dependent measurement. The coherence length based on this gap size is $\xi = \hbar v_F/\pi\Delta \approx$ 1.2 $\mu$m, where $\hbar$ is the reduced Planck constant and $v_F\approx$ 1 $\times$ 10$^{6}$ m/s is the Fermi velocity of graphene. The MAR signatures are only observed along terminals 1 and 3, with distance $\sim$ 0.8 $\mu$m $< \xi$, whereas no MAR is observed along terminals 2 and 4, with distance $\sim$ 3 $\mu$m $> \xi$. This observation is consistent with phase-coherent Andreev processes where the superconducting phase keeps (loses) its coherence before being reflected multiple times between terminals 1 and 3 (2 and 4). 

\begin{figure}
\includegraphics{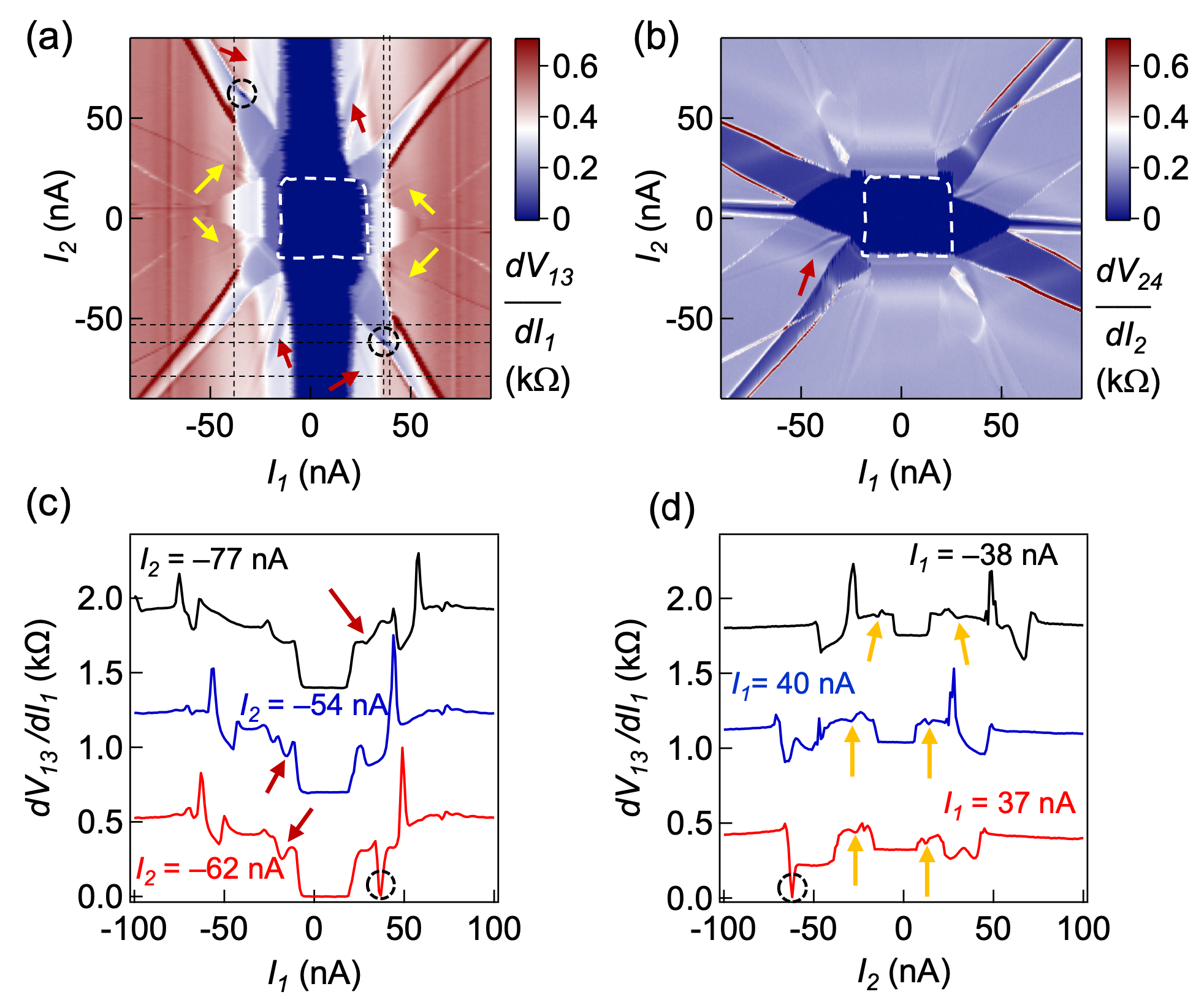}
\caption{\label{FIG2-01.png} (a-b) Color map of the differential resistance $dV_{13}/dI_{1}$ (a) and $dV_{24}/dI_{2}$ (b) versus $I_{1} =-I_{3}$ and $I_{2}=-I_{4}$ of device A at $V_g = 50$ V and $T = 12$ mK. The differential resistance is measured using a lock-in amplifier. In addition to the branches that have been labeled in Fig.~\ref{FIG1-01_NEW.png}(b), we observe re-entrant superconductivity (black circles), and narrow branches (marked by yellow and red arrows) along local minima of $dV_{13}/dI_{1}$. Dashed white contours outline the critical current contour corresponding to $V = 0$ V for all superconducting terminals. (c,d)  $dV_{13}/dI_{1}$ vs $I_{1}$ (c) and $dV_{13}/dI_{1}$ vs $I_{2}$ (d) along horizontal and vertical black dashed lines in panel (a), respectively. The local minima and re-entrant superconductivity are marked with arrows and circles as in (a). Curves are shifted vertically by $0.7$ k$\Omega$ for clarity.}
\end{figure}

To more closely examine the fine structure around the center region in Fig.~\ref{FIG1-01_NEW.png}(b), we plot $dV/dI$ over a smaller range of $I_{1}$ and $I_{2}$ in Fig.~\ref{FIG2-01.png}. Figs.~\ref{FIG2-01.png}(a) and (b) plot $dV_{13}/dI_{1}$ and $dV_{24}/dI_{2}$ versus $I_{1}$ and $I_{2}$, respectively. The critical current contour (CCC) is indicated with white dashed lines on these maps. We observe several resonant features (narrow branches marked by red and yellow arrows) corresponding to local minima of $dV/dI$. To better highlight these minima, Figs.~\ref{FIG2-01.png}(c) and (d) depict cuts along the dashed horizontal and vertical lines in Fig.~\ref{FIG2-01.png}(a), respectively. 
  
\begin{figure}
\includegraphics[scale=.075]{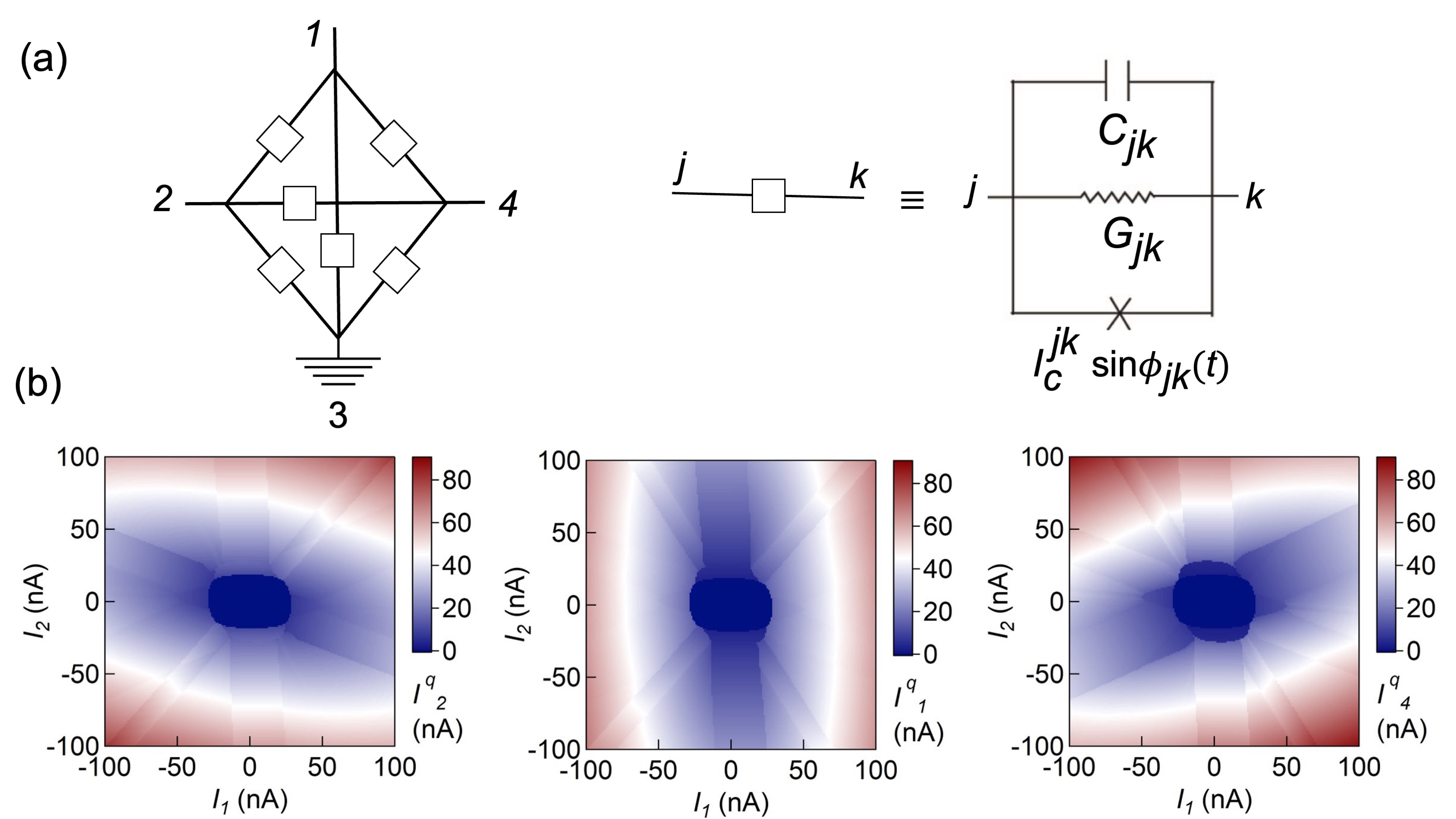}
\caption{\label{qcurrents} (a) Schematic of the circuit-network of coupled RCSJs utilized to simulate a four-terminal JJ in which the third terminal is grounded. The links between terminals, e.g., $j$ and $k$, are characterized by the sinusoidal CPR $I_c^{jk}\sin \phi_{jk}(t)$, a shunted conductance $G_{jk}$, and capacitance $C_{jk}$. (b) Calculated quasiparticle current $I^q_j$ as a function of the input currents obtained from the coupled RCSJ model. The boundary of the region where $I^q_j=0$ determines the CCC.}
\end{figure}

To elucidate the underlying origin of the observed features, we now consider a circuit-network model of coupled RCSJs (Fig.~\ref{qcurrents}(a)). The RCSJ model represents the individual junctions by a two-fluid system in which the total junction current is the sum of a dissipative quasiparticle current $i_{jk}^q(t)$ and a pair current $i_{jk}^p(t)$~\cite{mccumber1968effect}. The quasiparticle current is due to a finite voltage $V_{jk}(t)$ across the junction that exceeds the superconducting gap, $i_{jk}^q(t)=G_{jk}V_{jk}(t)$, where $G_{jk}$ is a constant phenomenological conductance tensor. The pair current is given by the diffusive CPR $i_{jk}^p(t)=I^{jk}_c\sin(\phi_{jk}(t))$, where $I^{jk}_c$ is the critical current and $\phi_{jk}(t)\equiv\phi_j(t)-\phi_k(t)$ is the gauge-invariant phase difference satisfying the Josephson equation $d\phi_{jk}(t)/dt=(2e/\hbar)V_{jk}(t)$. Here, we assume a diffusive CPR for the junctions. However, we note that the CPR for ballistic, diffusive, or $\varphi_0$ junctions do not change our overall conclusions (see Secs.~\ref{ballistic} and \ref{phi_0} for more details). Subsequently, we assume that the junctions are characterized by the presence of a shunt capacitance $C_{jk}$, which is the characteristic of circuits involving Josephson junctions in a wide family of weak links~\cite{mccumber1968tunneling}. Imposing current 
conservation (Kirchhoff's current law) at the terminal $j$ yields

\begin{equation}\label{rcsj}
    I_j=\sum_{k}(i_{jk}^p+i_{jk}^q+C_{jk}\frac{dV_{jk}(t)}{dt}).
\end{equation}

Eq.~\ref{rcsj} results in three coupled differential equations, that may be solved for the relevant junction phases $(\phi_2(t),\phi_1(t),\phi_4(t))$ assuming that one of the terminals is grounded, i.e., $\phi_3(t)=0$ as shown in Fig.~\ref{qcurrents}(a). The dc voltages, relative to the grounded terminal, are obtained by taking the time average as $  \langle V_{j3}(t)\rangle=(\hbar/2e)\langle d\phi_{j}(t)/dt\rangle\equiv V_j$.

To separate out the contributions of quasiparticle and pair currents to the total current flowing between different terminal pairs, we take the time average of Eq.~\ref{rcsj}. We note that any non-zero voltage between terminals $j$ and $k$ results in a non-zero quasiparticle current: $I^q_{jk}\equiv G_{jk}(V_j-V_k)$. We define the CCC as the region wherein the current is solely carried by the pair contribution and, accordingly, the quasiparticle contribution is zero. Fig.~\ref{qcurrents}(b) shows $I^q_j\equiv \big(\sum_{k}(I_{jk}^q)^2\big)^{1/2}$ as a function of the input currents. In the central region, dark blue area in Fig.~\ref{qcurrents}(b), $I^q_j=0$ and all the terminals are at zero voltage.  

\begin{figure}
\includegraphics[scale=.075]{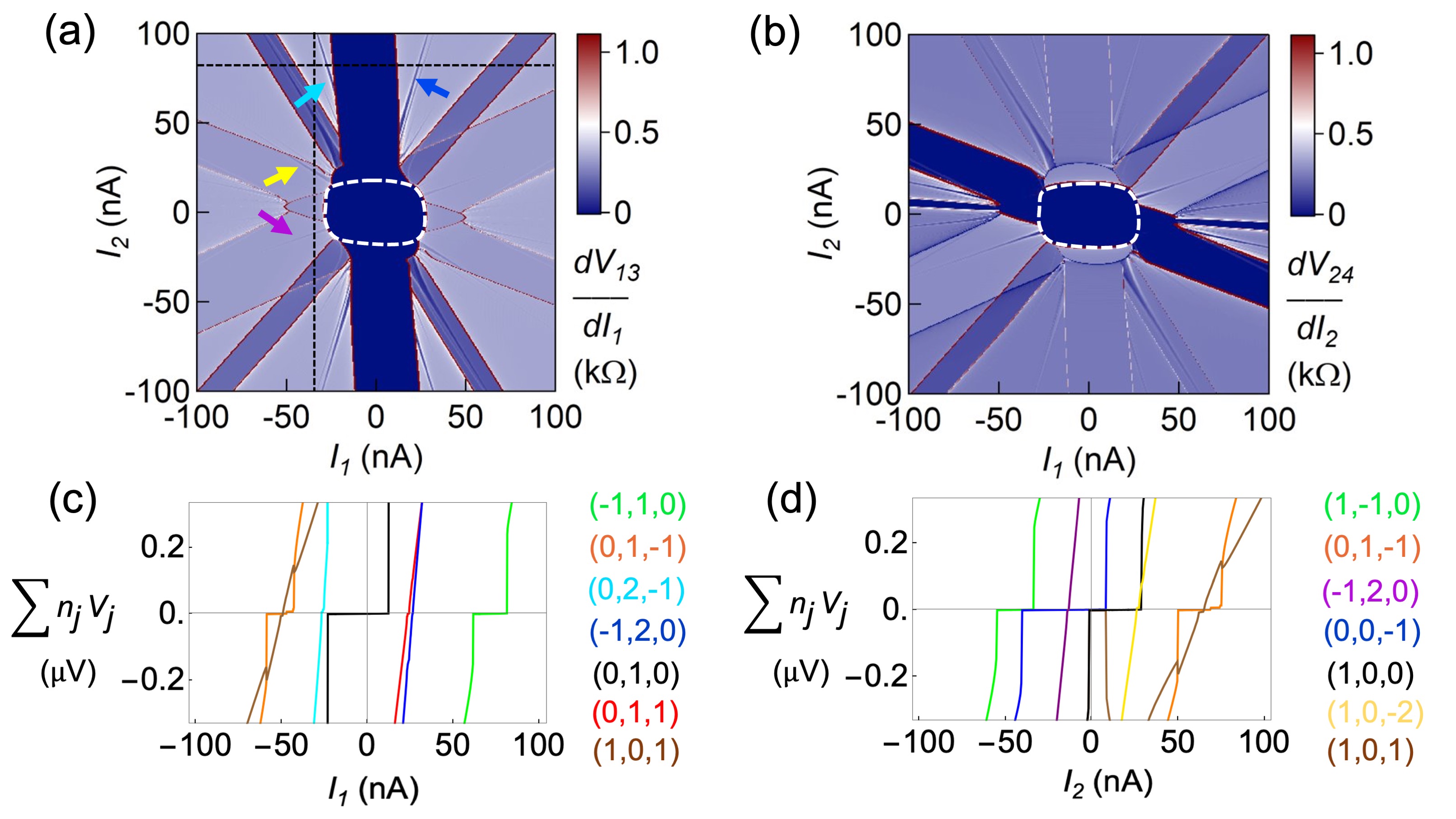}
\caption{\label{MTJJ} Theoretical simulation of differential resistance $dV_{13}/dI_1$ (a) and $dV_{24}/dI_2$ (b) versus $I_{1}$
and $I_2= - I_4$ obtained from the coupled RCSJ model. 
The white dashed contours show the CCC, which encloses the region in which $I^q=0$. (c) $\sum_jn_jV_j$ as a function of $I_1$ 
for a line cut at $I_2=80$ nA, along the horizontal dashed line in panel (a). (d) $\sum_jn_jV_j$ as a function of $I_2$ for a 
line cut at $I_1=-40$ nA, along the vertical dashed line in panel (a). Each branch that is crossed by the black dashed line in panel (a) corresponds to a unique
triplet $(n_2,n_1,n_4)$ satisfying $\sum_jn_jV_j=0$. Only the narrow branches, indicated with the small arrows, correspond to higher-order harmonics.}
\end{figure}

Fig.~\ref{MTJJ} depicts the calculated differential resistance $d(V_{j}-V_{k})/dI_j\equiv dV_{jk}/dI_j$ as a function of the input currents, $I_1$ and $I_2$. We observe that while the RCSJ model successfully reproduces all the major branches including those marked by the dashed lines in Fig.~\ref{FIG1-01_NEW.png}(b) and arrows in Fig.~\ref{FIG2-01.png}(a), it does not capture the phase-coherent processes such as the MARs (vertical lines in Fig.~\ref{FIG1-01_NEW.png}(b)). We further note that the observed branches in the $dV/dI$ maps exhibit a radial inversion symmetry, which is imposed by the Onsager reciprocity in the absence of a magnetic field. As a result, these branches can be identified by a unique triplet $(n_2,n_1,n_4)$ satisfying

\begin{equation}\label{har}
    \sum_jn_jV_{j}=0,
\end{equation}

where $n_2$, $n_1$, and $n_4$ are integers. Figs.~\ref{MTJJ}(c) and (d) highlight such triplets for branches crossing the black dashed lines in Fig.~\ref{MTJJ}(a).

To provide an analytical picture for the overall CPR of our four-terminal JJ, we consider all the terminals are at zero voltage and $I_4=0$. We note that these assumptions are made to simplify the following analytical derivations and will not affect our overall conclusions (see Appendix~\ref{app:float} for more details). The energy $F$ of the system is given by

\begin{equation}\label{Energy}
    F=-\frac{\hbar}{2e}\big(I_1\phi_1+I_2\phi_2+\sum_{j<k}I_c^{jk} \cos\phi_{jk}\big), \,\,\,\,\,\, \phi_3=0,
\end{equation}
We obtain $\phi_4$ by minimizing $F$ with respect to $\phi_4$ for fixed $(\phi_2,\phi_1)$ as

\begin{equation}\label{pj}
    \phi_4=\arctan\frac{\sum_{k\neq 4}I_c^{4k}\sin\phi_{k}}{\sum_{k\neq 4}I_c^{4k}\cos\phi_{k}}, \,\,\,\,\,\, \phi_3=0.
\end{equation}
We can express the current flowing from terminal 4 to ground as
\begin{equation}
    i^p_{43} = I_c^{43}\sin\phi_4.
\end{equation}
Generically, since $\phi_4$ is a $2\pi$-periodic odd function of $(\phi_2,\phi_1)$, $i^p_{43}$ can be expanded as a Fourier series~\cite{melo2022multiplet}:
\begin{equation}\label{gJ}
    i_{43}^p=\sum_{n_2,n_1}I_{n_2,n_1}\sin(n_2\phi_{2}+n_1\phi_1),
\end{equation}
where $(n_2,n_1)$ are integers and $I_{n_2,n_1}$ is the amplitude of the $(n_2,n_1)$ harmonic. We note that in a general case of $I_4\neq 0$, a triplet $(n_2,n_1,n_4)$ may emerge (see Figs.~\ref{MTJJ}(c,d)).

According to Eq.~\ref{gJ}, a dc pair current flows if $I_{n_2,n_1}$ is non-zero and $n_2\phi_2+n_1\phi_1$ is constant~\cite{melo2022multiplet,PhysRevB.102.064510}. Inside the CCC, the pair current is non-zero because the phases are time-independent, leading to a zero voltage on each terminal ($V \propto d\phi/dt$). We note that outside of the CCC, voltages are non-zero and Eq.~\ref{Energy} is no longer valid. However, in a non-equilibrium case in which both quasiparticle and pair currents may flow, one may utilize the non-equilibrium Green’s 
functions method ~\cite{PhysRevB.87.214501} to recover an analogous current-phase relation to Eq.~\ref{gJ}.
Our numerical analysis suggests that outside CCC, a pair current may still emerge on the resonant branches where pairwise combinations of $(n_2,n_1)$ satisfies $n_2\phi_2(t)+n_1\phi_1(t)=\text{const.}$

We point out that starting from a semi-classical RCSJ model, in which terminals are pairwise coupled, the circuit-network model 
may generate a non-local transfer of Cooper pairs between two terminals. For example, Eq.~\ref{gJ} suggests that $n_1$ and $n_2$ Cooper 
pairs are respectively transferred from ground to terminals 1 and 2 through terminal 4~\cite{Melin_2014}. Therefore, it is challenging to separate non-local phase-coherent 
Andreev reflection 
processes from the semi-classical circuit-network effects as both processes may result in similar macroscopic transport observables. 
More sophisticated measurements such as correlated noise spectroscopy \cite{melin2016gate,cohen2018nonlocal} provide additional information required to 
distinguish these processes.  
We finally note that the RCSJ model does not reproduce the re-entrant superconductivity, marked by black dashed circles in 
Fig.~\ref{FIG2-01.png}. Here, the re-entrant superconductivity refers to the case where superconductivity, i.e., $dV/dI$ = 0, 
re-emerges at a non-zero bias voltage. Understanding the origin of this phenomenon is a subject of future studies. 

\begin{figure}
\includegraphics{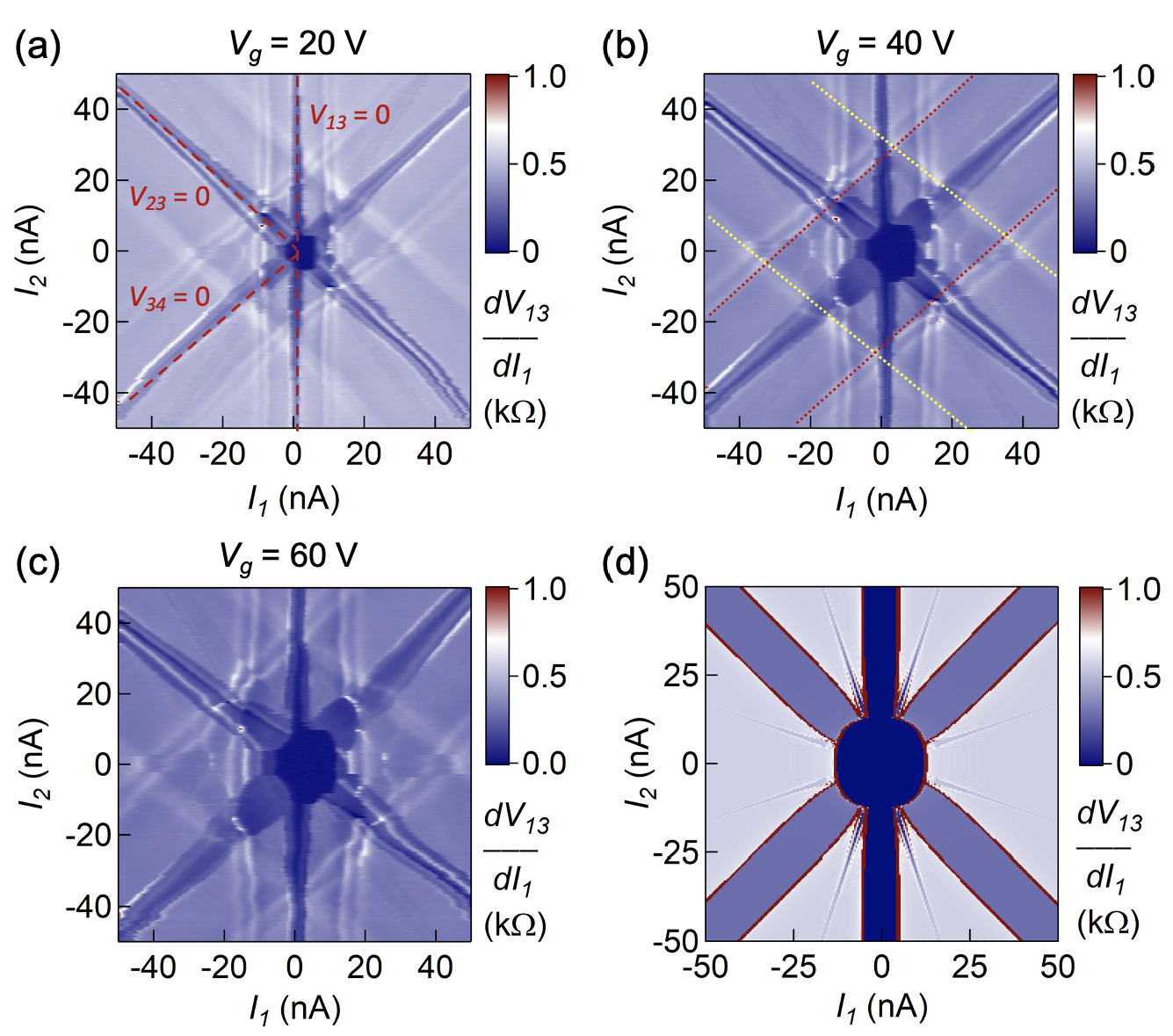}
\caption{\label{FIG4-new.png} (a-c) Color maps of the differential resistance ($dV_{13}/dI_1$) versus $I_{1}=-I_{3}$ and $I_{2}=-I_{4}$ of device B at three different gate voltages $V_g=20$ V (a), $40$ V (b), and $60$ V (c). Dashed red lines depict three superconducting branches corresponding to $V_{jk}$ = 0 between pairs of superconducting terminals as labeled. We also observe repeating local minima of differential resistance, marked by red and yellow dotted lines in (b), that are parallel to branches labeled in (a). The differential
resistance is measured using a lock-in amplifier. All the measurements are performed at $T = 12$ mK. (d) Theoretical simulation of the differential resistance ($dV_{13}/dI_1$) for a symmetric RCSJ configuration, i.e., $I_c^{jk}=5$ nA, and $G_{jk}=1/115$ $\Omega^{-1}$. 
} 
\end{figure}

\section{Transport in the symmetric device geometry}\label{symmetry}
We now focus on the symmetric device (device B), wherein the graphene region (1.3 $\mu$m in diameter) falls within the superconducting coherence length ($\xi \sim$ 1.2 $\mu$m). Figs.~\ref{FIG4-new.png}(a-c) show the $dV_{13}/dI_1$ maps versus $I_{1}$ and $I_{2}$ at three different gate voltages $V_g = 20$ V, $40$ V, and $60$ V, respectively. We observe zero-resistance branches in the $dV_{13}/dI_1$ map, corresponding to supercurrent flow between pairs of terminals, as indicated by red dashed lines in Fig.~\ref{FIG4-new.png}(a). Fig.~\ref{FIG4-new.png}(d) depicts the $dV_{13}/dI_1$ map calculated from the RCSJ model, where the same parameters are used for all the junctions. We observe that while the theoretical map captures the main features of the experimental data, e.g., those marked by dashed red lines in Fig.~\ref{FIG4-new.png}(a), it does not reproduce the repeating branches of local minima, e.g., those marked by dashed yellow and red lines in Fig.~\ref{FIG4-new.png}(b). We attribute the repeating branches to MARs among adjacent superconducting terminals \cite{pankratova2020multiterminal}. For example, the yellow and red dotted lines correspond to MARs for $e|V_{23}|=2\Delta/6$ and $e|V_{34}| = 2\Delta/9$, respectively. We also do not observe any signature of multiplet pairings (narrow branches) in device B. We note that the induced superconductivity (area of the CCC) is weaker (smaller) in device B compared to device A, likely due to lower contact transparency. Therefore, we are unable to resolve the multiplet signatures (narrow branches) within  experimental resolution. This is consistent with our observations that more multiplet branches appear in device A at $V_g = 50$ V (Fig.~\ref{FIG2-01.png}(b)) where the induced superconductivity is stronger compared to $V_g = -50$ V (see Appendix ~\ref{app:field} for more details). Finally, we observe that by decreasing $V_g$ from $60$ V to $20$ V in the electron-doped regime (note the Dirac point is at $V_g = -11.25$ V), the area of the CCC monotonically decreases. In contrast to a previous report \cite{pankratova2020multiterminal}, we observe that the global gate only influences the size of the CCC; we do not see any obvious change in the geometry of the CCC. This discrepancy may be related to the symmetric nature of the Fermi surface and lack of spin-orbit coupling in graphene compared to InAs.

\section{RCSJ model with ballistic junctions}\label{ballistic}
So far we have only considered diffusive transport in the RCSJ model. However, in graphene, it is important to consider ballistic limit to understand the evolution of the differential resistance and resonant features. To be more specific, we consider that transport is mainly facilitated by the Andreev bound states localised in the junction region. As a result, the energy of the Andreev bound states for the junction between terminals $j$ and $k$ is given by $\varepsilon_{n,jk}=\Delta\sqrt{1-T_{n,jk}\sin^2(\phi_{jk}/2)}$, where $T_{n,jk}$ is the transmission eigenvalue of the transport channel $n$. This results in the following CPR

\begin{equation}
    I_{jk}=-\frac{2e}{\hbar}\sum_n\frac{d\varepsilon_{n,jk}}{d\phi_j}.
\end{equation} 

\begin{figure}
\includegraphics[scale=.08]{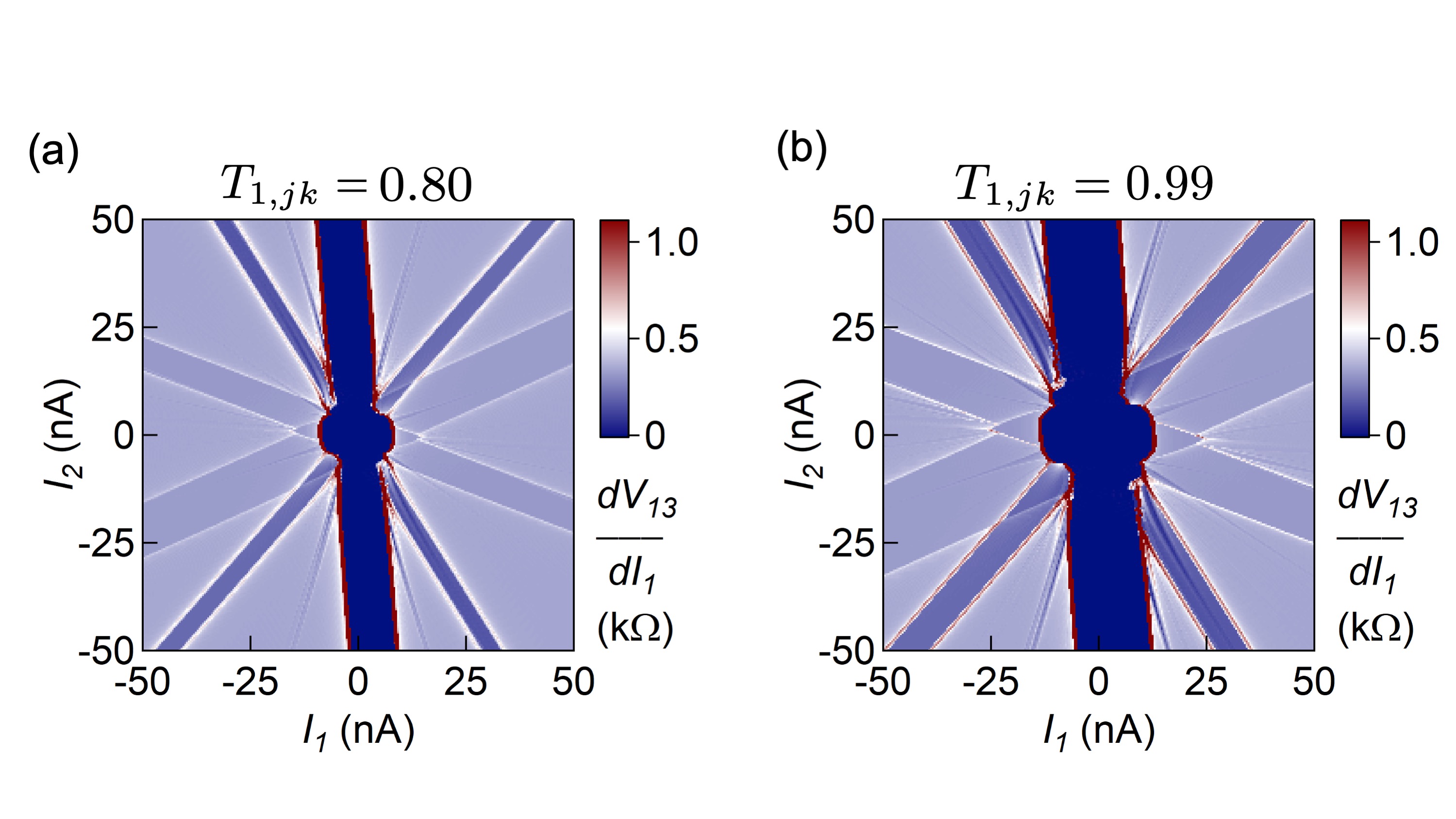}
\caption{\label{abs}  Theoretical simulation of differential resistance $dV_{13}/dI_1$ versus $I_{1}$ and $I_2= - I_4$ obtained 
from the coupled RCSJ model for a different junction transparency (a) $T_{1,jk}=0.80$, and (b) $T_{1,jk}=0.99$.}
\end{figure}

We note that with a strong elastic scattering at the junction region, i.e., $T_{n,jk}\ll1$, the sinusoidal CPR can be recovered. Therefore, we only focus on the clean regime $T_{n,jk}\sim1$ where the CPR deviates from the usual sinusoidal profile by developing a skewness. Fig.~\ref{abs} shows the simulated differential resistance $dV_{13}/dI_1$ versus $I_1$ and $I_2$ for $T_{1,jk}=0.8$ (a) and $T_{1,jk}=0.99$ (b) . We observe that the differential resistance maps are qualitatively similar to the one shown in Fig.~\ref{MTJJ}(a). This indicates that the resonant features in differential resistance maps are robust to the variation of the junction transparency.

\section{RCSJ model with $\varphi_0$-junctions}\label{phi_0}
\begin{figure}
\includegraphics[scale=.08]{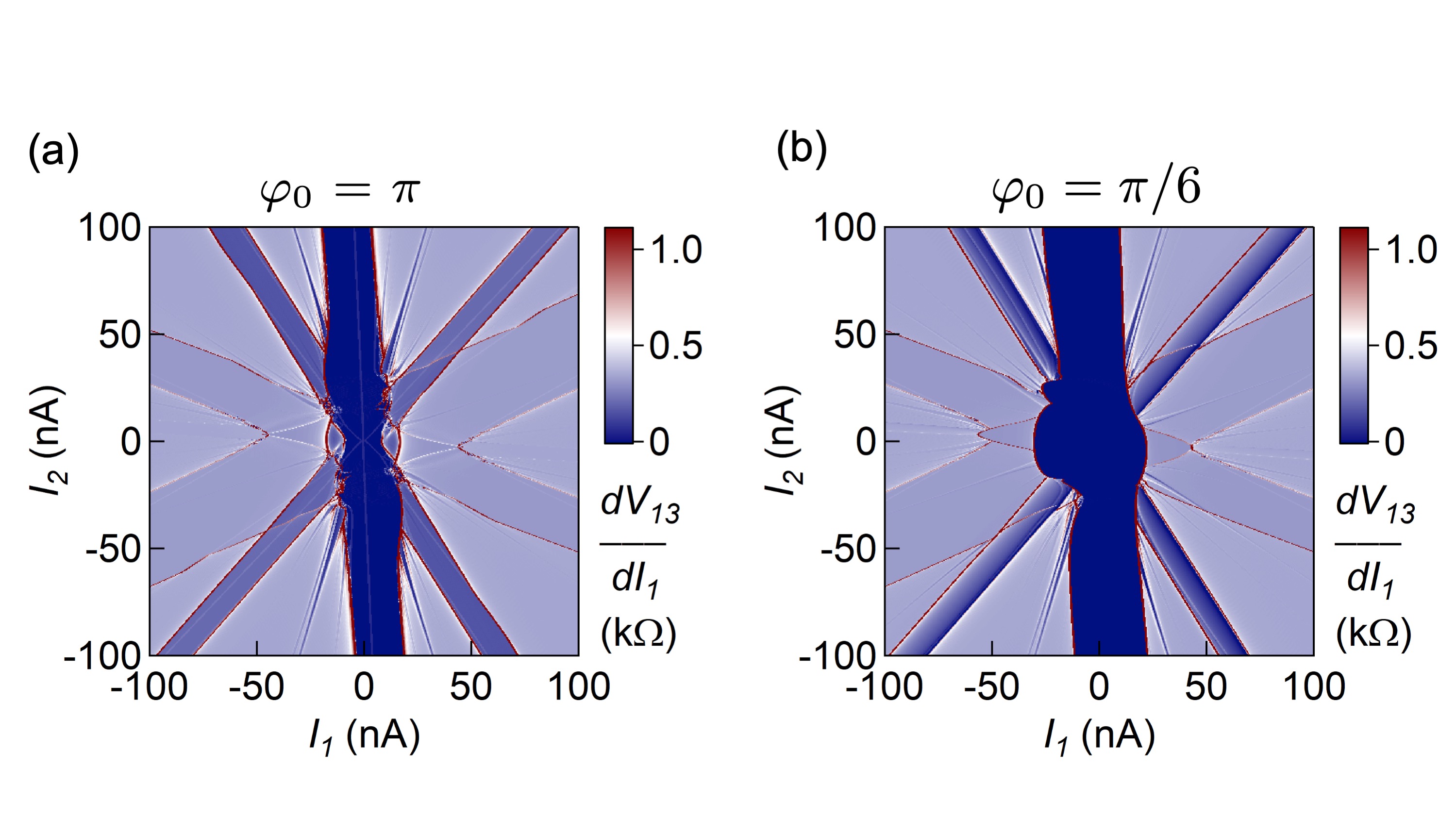}
\caption{\label{phiJ}  Theoretical simulation of differential resistance $dV_{13}/dI_1$ versus $I_{1}$ and $I_2= - I_4$ obtained 
from the coupled RCSJ model for $\varphi_0$-junctions with a CPR between terminals $j$ and $k$ as $i^p_{jk}=I_c^{jk} \sin(\phi_{jk}-\varphi_0)$. Panel (a) is for $\varphi_0=\pi$, and (b) is for $\varphi_0=\pi/6$.}
\end{figure}

The CPR depends on the Fermi surface geometry of the normal material and, concordantly, modifications to the host electronic band structure modify this CPR. More explicitly, a Josephson junction with Rashba spin-orbit coupling and magnetic order in the normal region may give rise to a sinusoidal CPR with a phase offset, $\sin(\phi-\varphi_0)$ ~\cite{buzdin2008direct}. Here, we consider the coupled RCSJ model with a modified CPR between junctions as 
$i^p_{jk}=I_c^{jk} \sin(\phi_{jk}-\varphi_0)$. Figs.~\ref{phiJ}(a) and (b) show the differential resistance $dV_{13}/dI_1$ maps versus $I_{1}$ and $I_2= - I_4$ obtained 
from this  RCSJ model for $\varphi_0 = \pi$ and $\pi/6$, respectively. We observe that the CCC is significantly modified by $\varphi_0$, whereas the resonant features outside of the CCC remain mainly independent of $\varphi_0$ and are qualitatively similar to those in Fig.~\ref{MTJJ}(a).  In general, the phase shift $\varphi_0$ may originate from various microscopic mechanisms such as broken inversion and time-reversal symmetries that affect the details of the Fermi surface in the normal region. 

\section{Conclusion}\label{conclusion}

In this work, we performed differential resistance measurements of symmetric and asymmetric four-terminal JJs. In addition to zero-resistance branches corresponding to supercurrent flow between pairs of superconducting terminals, we observed resonant features resembling multiplet Cooper pairings in the differential resistance maps. We observed that the size of the CCC monotonically increased with the increasing gate voltage due to the symmetric Fermi surface of graphene. We modeled our junctions using a network of coupled RCSJs to elucidate the experimental results. We theoretically investigated the contributions of quasiparticle and pair currents to the total current. Crucially, we found resonant features arising from circuit-network effects that mimic signatures of distinctly quantum processes such as multiplet pairings. Our calculations further demonstrated that the resonant features are insensitive to the exact form of the CPR. Our joint experimental and theoretical study paves the way toward using MTJJs as a materials-agnostic platform for engineering complex superconducting phases.

\begin{acknowledgments}
We acknowledge funding from the National Science Foundation (NSF) Innovation and Technology Ecosystems (No. 2040667). F.Z. and N.S. acknowledge support from the University of Chicago. G.J.C. acknowledges support from the ARAP program of the Office of the Secretary of Defense. M.J.G. and M.T.A. acknowledge funding from US ARO Grant W911NF-20-2-0151 and the NSF through the University of Illinois at Urbana-Champaign Materials Research Science and Engineering Center DMR-1720633. 

\end{acknowledgments}

\nocite{*}

\bibliography{apssamp}

\begin{appendix}
\section{RCSJ model parameters}\label{app:rcsj}
Here we provide the phenomelogical parameters used in our simulations. Dividing Eq.~\ref{rcsj} by $I_c=10 \,\text{nA}$ and defining the Josephson time constant as $\tau_J=\hbar/2e RI_c$ leads to

\begin{equation}
   \boldsymbol{I}=\boldsymbol{f}(\tau)+\mathcal{G}\cdot\frac{d\boldsymbol{\Phi}(\tau)}{d\tau}+\mathcal{C}\cdot\frac{d^2\boldsymbol{\Phi}(\tau)}{d\tau^2},
\end{equation}
where $\tau\equiv t/\tau_J$, and we set $R=1 \Omega$. Here, $ \boldsymbol{I}\equiv (I_2/I_c,I_1/I_c,I_4/I_c)^T$, $\boldsymbol{\Phi}(\tau)\equiv (\phi_2(\tau),\phi_1(\tau),\phi_4(\tau))^T$, and 

\begin{equation}
     \boldsymbol{f}(\tau)=\begin{pmatrix}
  I_{23}\sin\phi_{2}(\tau)+I_{21}\sin\phi_{21}(\tau)+I_{24}\sin\phi_{24}(\tau)\\
   I_{13}\sin\phi_1(\tau)+I_{12}\sin\phi_{12}(\tau)+I_{14}\sin\phi_{14}(\tau)\\
   I_{43}\sin\phi_{4}(\tau)+I_{42}\sin\phi_{42}(\tau)+I_{41}\sin\phi_{41}(\tau)
    \end{pmatrix}.
\end{equation}

The results presented in the main text are for the following dimensionless parameters 

\begin{equation}
    I^{jk}_c/I_c\equiv I_{jk}=
    \begin{pmatrix}
   I_{23} & I_{21} & I_{24}\\
   I_{13} & I_{12} & I_{14}\\
   I_{43} & I_{42} & I_{41}
    \end{pmatrix}=\begin{pmatrix}
   1.18 & 0.64 & 0.27\\
   2 &  0.64 & 0.64\\
   1.73 & 0.27 & 0.64
    \end{pmatrix},
\end{equation}

\begin{equation}
    \mathcal{G}_{jk}=
    \frac{1}{115}\begin{pmatrix}
   4.26 & -1 & -0.76\\
   -1 & 3.77 & -1\\
   -0.76& -1 & 6.26
    \end{pmatrix},
\end{equation}

\begin{equation}
    \mathcal{C}_{jk}=
    \frac{1}{1000}\begin{pmatrix}
   3 & -1 & -1\\
   -1 & 3 & -1\\
   -1 & -1 & 3
    \end{pmatrix}.
\end{equation}

As a result, the terminal dc voltage is given by $V_j=RI_c\langle d\phi_j/d\tau\rangle$. 

\renewcommand\thefigure{\thesection.\arabic{figure}}   
\setcounter{figure}{0}  

\section{Gate-voltage dependence of differential resistance}\label{app:field}

\begin{figure}[h]
\includegraphics{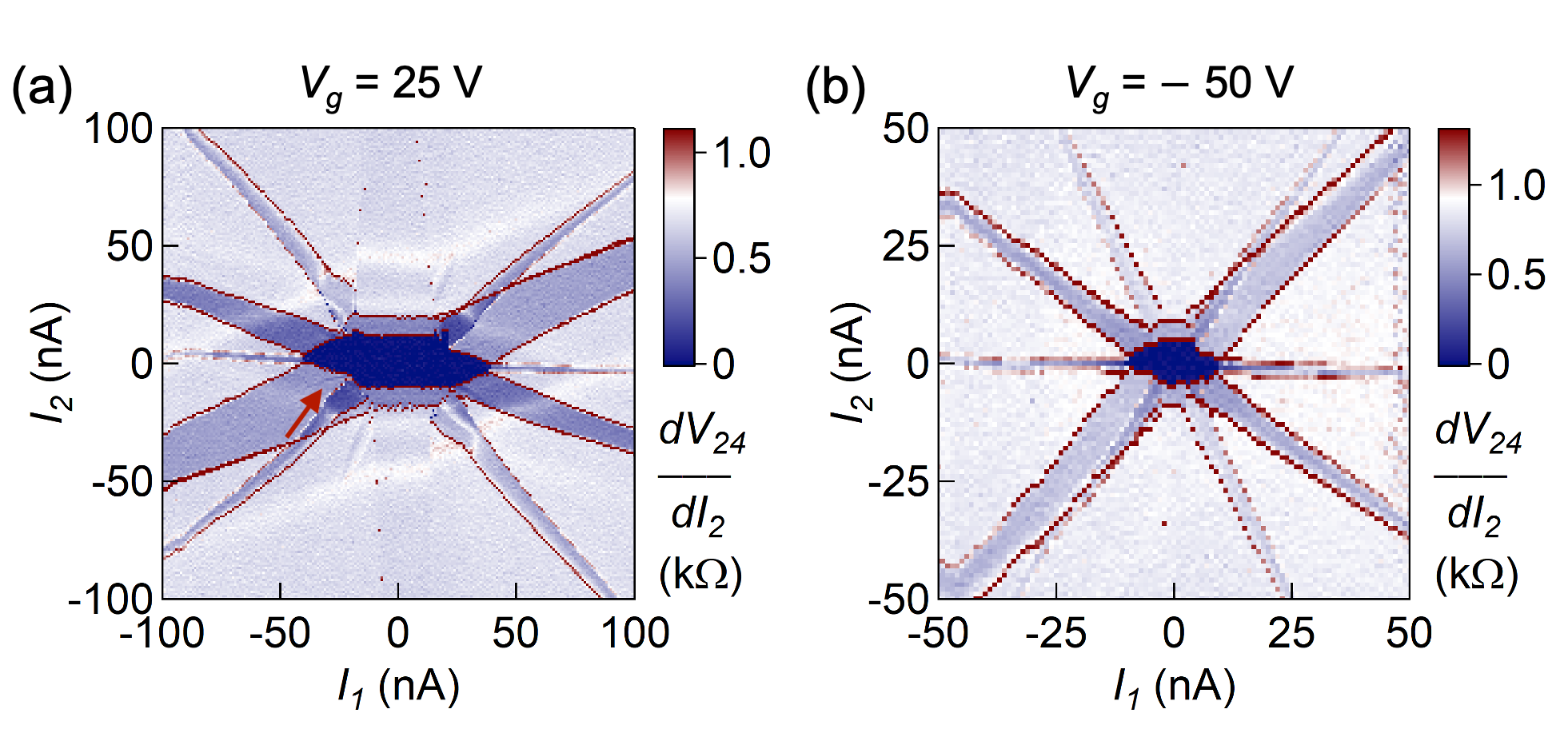}
\caption{\label{FIG5.png} Color map of differential resistance $dV_{24}/dI_2$ versus $I_{1}=- I_{3}$ and $I_{2}=-I_{4}$ in device A at (a) $V_g = 25$ V and (b) $V_g = -50$ V. The narrow branch (marked by red arrow) along local minima of $dV_{24}/dI_2$ is the same narrow branch as the one marked in Fig.~\ref{FIG2-01.png}(b). The differential resistance is obtained by taking a digital derivative of $V_{24}$ with respect to $I_2$. All measurements are performed at $T = 12$ mK.}
\end{figure}

Here, we present the gate dependence of differential resistance maps in device A. Fig.~\ref{FIG5.png} shows $dV_{24}/dI_2$ versus $I_{1} =-I_{3}$ and $I_{2} = - I_{4}$ for $V_g=25$ V (a) and $V_g= -50$ V (b). We observe that the major branches corresponding to supercurrent flow between adjacent terminals persist at different gate voltages. However, we note that the CCC size decreases by reducing the gate voltage from $50$ V to $25$ V, consistent with our observations in device B. Furthermore, we observe the multiplet branch (marked by red arrows in Fig.~\ref{FIG2-01.png}(b) and Fig.~\ref{FIG5.png}(a)) becomes smaller as the gate voltage changes from $50$ V to $25$ V, and eventually vanishes at $V_g = -50$ V. This may be related to the strength of the induced superconductivity that becomes weaker at $V_g = -50$ V, compared to $V_g = 50$~V. 

\section{Differential resistance maps for $I_4 = 0$}\label{app:float}

\begin{figure}
\includegraphics[scale=.29]{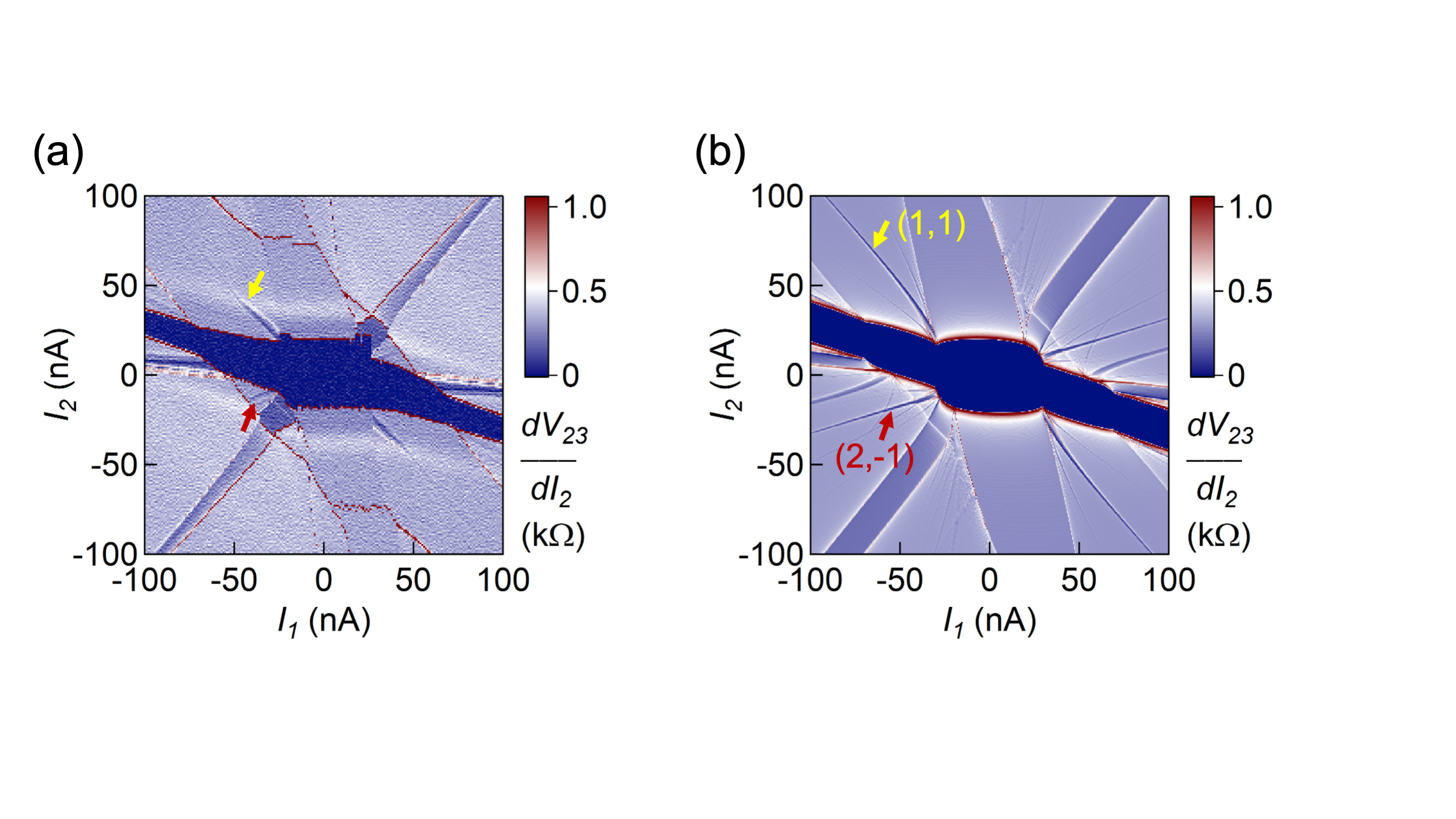}
\caption{\label{Floating} (a) Color map of differential resistance $dV_{23}/dI_2$ versus $I_{1}$ and $I_{2}$ in device A at $V_g = 50$ V. The differential resistance 
is obtained by taking a digital derivative of $V_{23}$ with respect to
$I_2$. (b) Simulated differential resistance $dV_{23}/dI_2$ versus $I_{1}$ and $I_{2}$. The narrow branches, marked by yellow and red arrows, along local minima of $dV_{23}/dI_2$ correspond
to $(n_2,n_1) = (1,1)$ and $(2,-1)$, respectively. The experimental and theoretical maps are obtained for a configuration where terminal 3 is grounded and $I_4 = 0$.}
\end{figure}

Here, we present our experimental and theoretical results for a measurement configuration where terminal 3 is grounded and $I_4=0$. Figures~\ref{Floating}(a) and (b) depict the experimental and simulated differential resistance plots, respectively. We observe resonant features corresponding to $(n_2,n_1) = (1,1)$ and $(2,-1)$ which are marked by yellow and red arrows in the figure, respectively. We note that only $(n_2,n_1)=(2,-1)$ represents a multiplet pairing. Our results demonstrate that multiplet pairings are robust and emerge regardless of the measurement configuration. 

\end{appendix}
\end{document}